\documentclass[12pt,preprint]{aastex}
\usepackage{natbib}
\usepackage{url}
\usepackage{amsmath}
\usepackage{appendix}
\usepackage{hyperref}
\begin{document}

\title{Two Thresholds for Globular Cluster Formation and their Dominance of Star Formation
in the Early-Universe}

\author{Bruce G. Elmegreen\altaffilmark{1}}

\altaffiltext{1}{IBM Research Division, T.J. Watson Research Center, 1101
Kitchawan Road, Yorktown Heights, NY 10598; bge@us.ibm.com}

\begin{abstract}
Young massive clusters (YMCs) are usually accompanied by lower-mass clusters
and unbound stars with a total mass equal to several tens times the mass of the
YMC. If this was also true when globular clusters (GCs) formed, then their
cosmic density implies that most star formation before redshift $\sim2$ made a
GC that lasted until today.  Star-forming regions had to change after this time
for the modern universe to be making very few YMCs. Here we consider the
conditions needed for the formation of a $\sim10^6\;M_\odot$ cluster. These
include a star formation rate inside each independent region that exceeds
$\sim1\;M_\odot$ yr$^{-1}$ to sample the cluster mass function up to such a
high mass, and a star formation rate per unit area of $\Sigma_{\rm
SFR}\sim1\;M_\odot$ kpc$^{-2}$ yr$^{-1}$ to get the required high gas surface
density from the Kennicutt-Schmidt relation, and therefore the required high
pressure from the weight of the gas. High pressures are implied by the virial
theorem at cluster densities. The ratio of these two quantities gives the area
of a GC-forming region, $\sim1$ kpc$^2$, and the young stellar mass converted
to a cloud mass gives the typical gas surface density of $500-1000\;M_\odot$
pc$^{-2}$. Observations of star-forming clumps in young galaxies are consistent
with these numbers, suggesting they formed today's GCs. Observations of the
cluster cut-off mass in local galaxies agree with the maximum mass calculated
from $\Sigma_{\rm SFR}$. Metal-poor stellar populations in local dwarf
irregular galaxies confirm the dominant role of GC formation in building their
young disks.
\end{abstract}
\keywords{globular clusters: general --- galaxies: clusters: general  ---
galaxies: star clusters: general  --- galaxies: star formation}

\section{Introduction}

The globular clusters (GCs) that surround the Milky Way and most other galaxies
formed when the universe was young, before a redshift of 1 to 2, in both
metal-poor dwarf galaxies  \citep{searle78,zinnecker88,freeman93} and less
metal-poor disk galaxies \citep{kravtsov05,shapiro10,tonini13,rossi15}, producing
the two populations we observe today \citep{harris91,brodie06,bastian18}. Many of
the dwarfs were captured and dispersed by larger host galaxies, leaving their GCs
in the host halos \citep{dacosta95,palma02,mackey04,gao07,van07,casetti09,
smith09, newberg09, mackey10}. Metal-rich globular clusters could have entered
galaxy halos when interactions stirred up their disks \citep{kruijssen15}. The
Sgr \citep{ibata94} and Canis Major \citep{martin04} dwarfs brought a dozen or
more GCs to the Milky Way \citep{forbes10,law10}. \cite{myeong18} suggest 8 GCs
came from another dwarf, and \cite{kruijssen18} suggest more came in earlier in a
dwarf they call ``Kraken.''

Tracing the origin of GCs is difficult because they were not bright enough when
they formed to observe directly with present techniques. Gravitational lensing
has revealed compact star-forming regions slightly more massive than GCs
\citep{vanzella17} and perhaps young GCs themselves \citep{johnson17,boylan18}.
Lyman alpha emitting galaxies \citep{finkelstein15,zheng17} may be a better
source for finding GCs directly, as these galaxies can be observed in deep
narrow-band Ly$\alpha$ surveys, their luminosities are consistent with star
formation rates expected in GC-forming regions, and their luminosity function at
low-mass gives the observed space density of today's metal-poor GCs
\citep{elmegreen12}. Dwarf galaxies in deep fields (``Little Blue Dots'') with
extremely high specific star formation rates, $\sim0.1$ Myr$^{-1}$ (star
formation rate per unit stellar mass), are another possible source for metal-poor
GCs \citep{elmegreen17a}.

Globular cluster formation seems to be an important if not {\it dominant} mode of
star formation in the early Universe. The space density of globular clusters
today follows from the product of the average number per unit luminosity of their
host galaxies and the number per unit volume of host galaxies with that
luminosity. \cite{pm00} derived 8 GCs Mpc$^{-3}$. For a $-2$ power law cluster
mass function \citep{port10}, the total cluster mass associated with a most
massive cluster of mass M$_{\rm max}$ is $M_{\rm max} \left(1+\ln(M_{\rm
max}/M_{\rm min})\right)$ \citep{elmegreen12} for minimum cluster mass $M_{\rm
min}$. With an approximately 25\% fraction of new stellar mass going into
clusters \citep{chandar17}, the total stellar mass associated with a
$10^6\;M_\odot$ cluster is $5\times10^7\;M_\odot$, a factor of 50 mass multiplier
\citep[weakly dependent on $M_{\rm min}$, which was assumed to be $100\;M_\odot$
from][]{lada03}. If these $10^6\;M_\odot$ clusters are the 8 Mpc$^{-3}$ GCs
around today, then the stellar density from their formation is
$4\times10^8\;M_\odot$ Mpc$^{-3}$. This is the same as the average co-moving
stellar density in the universe at a redshift of around 1 in the compilation by
\cite{madau14}.  \cite{boylan18} similarly derive a major contribution by GCs to
the high-redshift galaxy luminosity function at the low end. If the space density
of metal-poor GCs is $\sim2$ Mpc$^{-3}$, as estimated by \cite{boylan17} using
the correlation between GCs and dark matter halos, then the associated stellar
density is about half the above value, considering also the metal-rich GCs that
formed later. Evidently, a high fraction of star-forming regions included at
least one globular cluster before about half the age of the universe.

A similar calculation illustrates the importance of GC formation to the universal
star formation rate at early times. If we consider that the duration of the star
formation event which made a GC was $\Delta t$, and the redshift range over which
they formed was from $z=7$ to 2, a period of 2.6 Gyr, then the average star
formation rate from GC-forming events with $M_{\rm max}\sim10^6\;M_\odot$ is
($5\times10^7\;M_\odot/\Delta t)\times(\Delta t/2.6 \;{\rm Gyr})\times(8\;{\rm
Mpc}^{-3})$, where the second term is the fraction of time each event occupies in
the GC-forming era. The resultant average associated star formation rate is
$0.15\;M_\odot$ yr$^{-1}$ Mpc$^{-3}$, which is the same as the peak value of the
co-moving cosmic star formation rate density in the universe that occurs at
redshift $\sim2$ \citep{madau14}.

These two calculations suggest that the early universe was filled with
star-forming regions containing young massive clusters that ended up as today's
GCs. There was little additional star formation making smaller clusters or OB
associations that dispersed, aside from the smaller clusters and associations
directly connected with the GCs.  Today the mass density of GC stars is their
space density times their average mass of $\sim2\times10^5\;M\odot$
\citep{harris01}, or $1.6\times10^6\;M_\odot$ Mpc$^{-3}$. This is 0.26\% of the
current total stellar mass density from \cite{madau14}, a factor of $\sim1/380$.
The first factor of 50 in this 380 is presumably from scattered stars that
formed with the GCs, and then the remaining factor of $\sim8$ would be from
stars that formed recently without making massive clusters.

We can also learn about early star formation in dwarf galaxies from their old
stellar populations. WLM has an old metal-poor GC whose luminosity suggests it
contains $\sim10^6\;M_\odot$ of stars \citep{elmegreen12}. The mass of other
stars at the same age and metallicity is comparable to this
\citep{leaman12,larsen14}, suggesting the GC was a high fraction of the total WLM
mass when it formed \citep{elmegreen12}. Similarly, the GC mass in the Fornax
dwarf spheroidal was $\sim0.25$ of the galaxy stellar mass when the GC formed
\citep{larsen12}. For IKN and NGC 147, these fractions are $\sim0.5$ and $0.06$,
respectively \citep{larsen14,larsen18}. These are enormously high fractions,
especially considering the additional cluster mass that might have been present
in the full cluster mass function.  Perhaps the clustered fraction of stars at
birth was larger than the 25\% assumed above, even close to 100\% as suggested by
\cite{messa18} for regions with high star formation rate densities. In either
case, the star formation events that formed the observed GCs also made most of
the stellar disks at the same time. These were {\it whole-galaxy starbursts}.

The specific frequency of GCs in local dwarf galaxies also suggests nearly the
whole galaxy was involved in the same starburst. The number of GCs per
$10^9\;M_\odot$ of stars is $\sim100$ for a stellar mass today of
$3\times10^8\;M_\odot$ \citep{zaritsky16}. Multiplying the number of GCs by their
current average mass, $\sim2\times10^5\;M_\odot$, gives a current mass fraction
of 2\%.  At 10\% of the age of the universe these GCs were there but the stellar
mass was only about one-tenth as much, making the GC mass fraction $\sim20$\% in
the dwarf galaxies. For more massive initial GCs, i.e., considering evaporation
and an escaped population of first-generation stars that contributed p-processed
elements to approximately half of the remaining stars \citep{bastian18}, and
considering also the accompanying star formation in the same burst, the GC mass
fraction gets close to $\sim100$\%.

The Milky Way has a relatively low total mass of stellar streams and halo stars
accompanying its GCs.  Observations by \cite{martell16} of Milky Way halo stars
with metallicities matching those of the second generation in GCs (e.g., high
Nitrogen) suggest only an equivalent mass of these stars in the halo.  This
equivalence suggests GC evaporation halved their initial mass \citep[see
also][]{vesperini98}. More important for the present discussion are the stars
accompanying the first generation in each GC, which outnumber the GC stars by a
factor of $\sim20$ in the halo \citep{martell17}. These stars presumably come
from the environment in which the GCs formed, including those in the GC-forming
event prior to self-enrichment or some other process \citep{bastian18}. While
this factor of 20 is more than enough to allow the first generation stars in each
GC to enrich the second generation and mostly escape into the halo
\citep{bastian18}, it is not enough to carry along with each GC a significant
mass of other stars in an associated dwarf galaxy. \cite{sollima17} inferred an
even higher fractional mass loss from Milky Way GCs using the mass distribution
functions of remaining stars, whose shape correlates with the mass loss fraction.
They suggest $2\times10^8\;M_\odot$ of stars escaped from GCs, compared to
$\sim10^9\;M_\odot$ in the halo \citep{morrison93,bell08,deason11}. These escaped
stars are so numerous that they leave little room for additional stellar mass in
accompanying dwarf galaxies.  This requires that most of the dwarf galaxies that
brought in metal-poor GCs were accreted so early that they were still dominated
by their GC-forming population.

Local dwarf galaxies rarely form clusters as massive as a globular cluster
\citep{billett02,larsen09} and they never have nearly all of their stellar mass
resulting from a single burst. NGC 1569 \citep{stil98} and NGC 5253
\citep{lopez12,turner15,miura15} have YMCs, but these clusters contain only 0.1\%
of the galaxy stellar masses \citep{demarchi97,johnson12, sabbi18,calzetti15}.
The origin of these YMCs may be related to impacting gas streams, which are
present in both cases. Most local starburst dwarfs have lopsided HI
\citep{lelli14}. A major gas impact is suspected also in the dwarf ``tadpole''
galaxy Kiso 5639, which has a molecular cloud at one end with a gas mass
comparable to the total stellar mass in the whole disk, and is currently forming
14 young star clusters more massive than $10^4\;M_\odot$
\citep{elmegreen16,elmegreen18}.  Mrk 930 is another starburst dwarf with a high
formation efficiency for clusters \citep{adamo11} and NGC 1705 has a YMC with no
evidence for an impact or merger. YMC formation in major mergers of large
galaxies is more common \citep{whitmore10}.

Clearly there was a transition at around redshift 1 to 2 from conditions that
made $10^6\;M_\odot$ clusters in most star-forming regions of the early universe
to present-day conditions that rarely make them except perhaps in major mergers.
We investigate this transition here, suggesting it is almost entirely the result
of a higher gas surface density at that time. Observations show this high surface
density directly \citep[e.g.,][]{tacconi18,cava18}, so the result may not be
surprising. There are few observation yet for the origin of this high surface
density, but the usual explanations of enhanced cold accretion
\citep[e.g.,][]{inoue16} and galaxy mergers \citep[e.g.,][]{yozin12,kim18}, are
both likely contributors.

An implicit assumption in this paper is that the basic star formation processes
at the redshifts where GCs formed were the same as they are in galaxies today.
These processes involve gaseous gravity and cooling, as mitigated and partitioned
by hydromagnetic turbulence and star formation feedback
\citep{mckee07,krumholz18}. As a result, the cluster mass function at birth and
the concurrent formation of clustered and non-clustered stars are probably
similar then and now too. Other possible assumptions such as a top-heavy cluster
mass function at high redshift are not necessary in the present model.

In what follows, Section \ref{minimum} derives the minimum star formation rate to
sample the cluster mass function up to a GC-forming mass of $\sim10^6\;M_\odot$.
The result, $\sim1\;M_\odot$ yr$^{-1}$, is commonly observed in today's galaxies
but typically spread out over a disk spanning many kpc$^2$, which corresponds to
too low an areal rate and too low a pressure to make a GC in any local region.
This leads to a second condition in Section \ref{sect:mcluster}, which is that the
surface density of star formation has to exceed $\sim1\;M_\odot$ pc$^{-2}$
Myr$^{-1}$ for the local pressure to compact the required amount of mass up to the
density of a star cluster.  These two relations suggest that $\sim1$ kpc is the
characteristic size of a GC-forming region and the minimum size of a GC-forming
galaxy (on average). The second condition also corresponds to the observed cut-off
mass in the local cluster mass function. This cut-off mass is predicted to be
difficult to observe unless the product of the surface area and the duration of
star formation is fairly large, because for a small value of this combined
parameter, the maximum mass that is likely to be observed from the sample size is
only comparable to or smaller than the cut-off mass (Sect. \ref{compare}). A brief
comparison of these two thresholds to observations of star formation at high
redshift is in Section \ref{clump}. This comparison implies that approximately
half of the star-formation in giant clumps has the required local star formation
rate and density to make a $10^6\;M_\odot$ cluster, which supports the above
suggestion that GCs and their associated stars were pervasive in the early
universe.

\section{Minimum star formation rate}
\label{minimum}

\begin{figure*}
\epsscale{.7}
\plotone{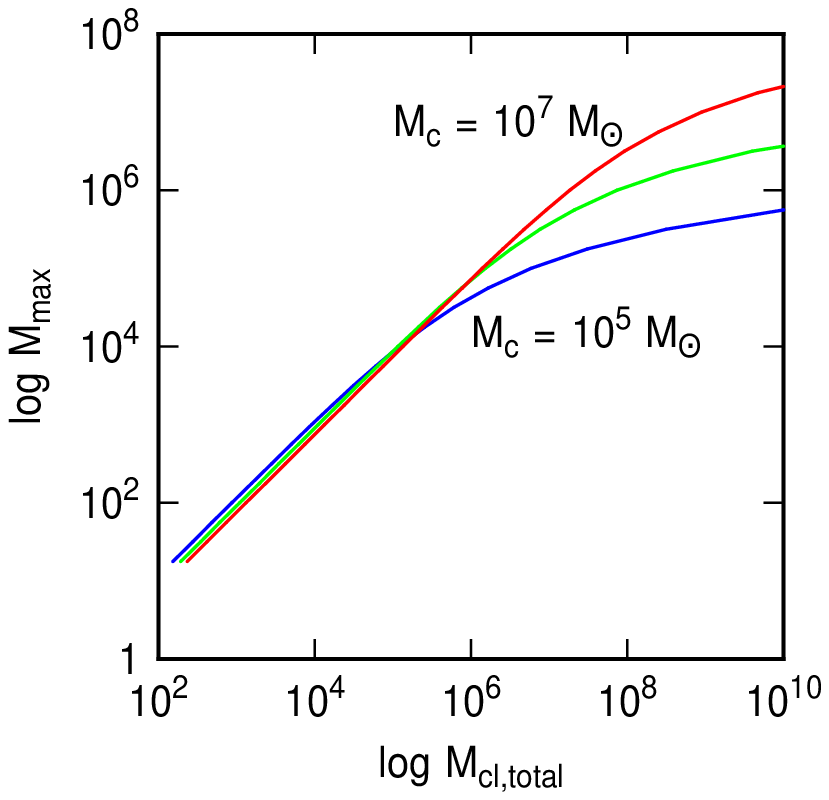}
\caption{The maximum likely mass of a star cluster as a function of the
total mass in clustered stars for three Schechter mass functions with different
cut-off masses. For the assumed $-2$ power law in the mass function,
the maximum likely cluster mass increases linearly with the total clustered mass
up to the cut-off mass, and then the total mass has to increase more rapidly to
get much of an increase in the maximum mass.}
\label{heidelberg18_large_total}
\end{figure*}

An important condition for the formation of a YMC is a high star formation rate,
so the cluster mass function, which is presumably from a random distribution of
clump masses in a turbulent, self-gravitating gas, can be sampled far out into the
high-mass tail where turbulent structures are rare. Assuming this mass function is
a Schechter function with a power law slope of $-2$ and a cutoff mass $M_{\rm c}$
\citep{gieles06a,gieles06b,jordan07,bastian12,adamo15}, i.e.,
\begin{equation}
dn(M)/dM=n_0M^{-2}\exp(-M/M_{\rm c}),
\label{schech}
\end{equation}
the maximum likely mass $M_{\rm max}$ is given by the condition that there is one cluster at or
larger than that mass,
\begin{equation}
\int_{M_{\rm max}}^{\infty}n(M)dM=1,
\end{equation}
which defines $n_0(M_{\rm max})$ . The total mass in all clusters is then
\begin{equation}
M_{\rm cl,total}(M_{\rm max})={{\int_{M_{\rm min}}^{\infty}M(dn[M]/dM)dM} \over
{\int_{M_{\rm max}}^{\infty}(dn[M]/dM)dM}}
\label{sum}
\end{equation}
where $M_{\rm min}$ is the minimum cluster mass. This equation can be solved to
give the maximum likely cluster mass as a function of the total cluster mass for
each value of $M_{\rm c}$.  Figure \ref{heidelberg18_large_total} shows the result
for the function given by equation (\ref{schech}). For $M_{\rm
c}\sim10^5\;M_\odot$ as in local galaxies
\citep{gieles06a,adamo15,adamo17,messa17,johnson17}, the total cluster mass in the
starburst has to be comparable to the mass of stars in a Milky Way size galaxy
before a $\sim10^6\;M_\odot$ cluster is expected, and this is not realistic. The
formation of a young GC with this mass in a dwarf galaxy requires a high M$_{\rm
c}\geq10^7\;M_\odot$, so the mass function up to the GC mass is effectively a pure
power law. Then the expression for total cluster mass is integrated to give
\begin{equation}
M_{\rm cl,total}=M_{\rm max}\left(1+\ln\left[{{M_{\rm max}}\over{M_{\rm min}}}\right]\right).
\label{mtot}
\end{equation}
(In the numerator of equation \ref{sum}, the partial integral over mass from
$M_{\rm max}$ to infinity was set equal to M$_{\rm max}$ to avoid a divergence in
the logarithm of infinity). If a fraction $\Gamma$ of stars form in clusters
according to this mass function and the rest are dispersed, then the total stellar
mass in the starburst event is $M_{\rm cl,total}/\Gamma$.

Equation \ref{mtot} suggests that the total young stellar mass associated with a
$\sim10^6\;M_\odot$ cluster is $5\times10^7\;M_\odot$ for $\Gamma\sim0.25$
\citep{adamo11,chandar17} and $M_{\rm min}\sim100\;M_\odot$. This mass should be
divided by the duration of star formation to get the average star formation
rate. The duration of star formation, $\Delta t$, is typically a crossing time
in a gravitating, turbulent region. For a giant clump in a disk galaxy or a
whole dwarf galaxy, the physical size might be $\sim1$ kpc and the velocity
dispersion $\sim20$ km s$^{-1}$, giving a characteristic timescale for the
starburst equal to $\Delta t\sim50$ Myr. Then the star formation rate is
$\sim1\;M_\odot$ yr$^{-1}$. The same duration results if we consider a total
efficiency of star formation equal to $2.5$\% and a gas consumption time of
$\sim2$ Gyr \citep[e.g.,][]{krumholz07,bigiel08}.

The uncertainty in $\Gamma$ at high and low redshift and the lack of observations
of the cluster mass function at high redshift limit the precise application of
this minimum star formation rate. However, variations in $\Gamma$ are likely to
be only a factor of a few, since it cannot be larger than 1, and the power law in
the cluster mass function seems to be a fundamental property of hierarchical
fragmentation \citep{fleck96,elmegreen97}, so variations in that are also likely
to be small.  In what follows, we continue to use $\Gamma\sim0.25$ as
representative during the epoch of GC formation. It enters only in the minimum
star formation rate derived here and not the minimum star formation rate surface
density considered in Section \ref{sect:mcluster}.

The results suggest that a region of star formation the size of a dwarf galaxy or
a giant clump in a larger clumpy galaxy (Sect. \ref{clump}),forming stars at a
rate, $S$, of at least $\sim1\;M_\odot$ yr$^{-1}$ for $\sim50$ Myr, should form
at least one $10^6\;M_\odot$ cluster that can be a possible predecessor to a GC.
We write this result as
\begin{equation}
M_{\rm max,SOS}\sim10^6\;M_\odot \left({S\over{1\;M_\odot\;{\rm yr}^{-1}}}
\right)\left({{\Delta t}\over{50\;{\rm Myr}}}\right),
\label{sos}
\end{equation}
where `SOS' refers to the Size of Sample effect, which was the basis for deriving
this mass. There is also a small logarithmic dependence on this maximum mass (eq.
\ref{mtot}) that has been ignored in this equation (i.e., $1+\ln(M_{\rm
max}/M_{\rm min})$ was set equal to 12.5).

As an example, consider the two local dwarf starburst galaxies where this
magnitude of star formation is associated with YMCs. The starburst in NGC 1569
made three clusters of mass $3.9\times10^5\;M_\odot$, $4.4\times10^5\;M_\odot$,
and $2.3\times10^5\;M_\odot$ \citep[][see also De Marchi et al. 1997, Larsen et
al. 2011]{gilbert02}, and the star formation rate in that region is
$\sim1\;M_\odot$ yr$^{-1}$ depending on the assumed initial stellar mass function
\citep{greggio98}. Also for NGC 5253 with two clusters of mass
$7.5\times10^4\;M_\odot$ and $2.5\times10^5\;M_\odot$, the star formation rate is
$0.4\;M_\odot$ yr$^{-1}$ over a timescale of $\sim10$ Myr \citep{calzetti15}.

Statistical sampling suggests that a galaxy with 10 regions of star formation
having a rate of $0.1\;M_\odot$ yr$^{-1}$ each should have the same probability of
forming a $10^6\;M_\odot$ cluster as a single region with $10\times$ the rate.
This is not the case in fact because the maximum mass of a cluster depends on the
pressure through the $M_{\rm c}$ in the Schechter function, and the pressure
depends on the star formation rate per unit area, not just the total rate. This
additional dependence will be discussed in the next section. The statistical
sampling argument, in which the maximum cluster mass is proportional to the number
or total luminosity of clusters with the same age
\citep{larsen02,rand13,kruijssen12b,whitmore14}, should work in principle even if
each star-forming region is independent, because the large regions themselves
should have a range of masses that follow a power law function, as shown in
Section \ref{clump}. Galaxies with total star formation rates of several $M_\odot$
yr$^{-1}$ like the Milky Way and local spirals do not form $10^6\;M_\odot$
clusters because each independent region has too low a pressure. We discuss in
section \ref{compare} how the maximum cluster mass from pressure considerations
compares with the maximum cluster mass from the size-of-sample effect.

\section{Minimum star formation surface density}
\label{sect:mcluster}
\subsection{Minimum Pressure}
\label{sect:minpres}

Massive clusters need high pressure environments to produce the high stellar
densities and masses of the gravitating cloud cores in which they form
\citep{elmegreen97}, and to offset the effects of young stellar feedback
\citep{kruijssen12}.  The most direct way to see this is from the virial theorem,
which can be used to relate the mass of a self-gravitating cloud to the pressure
and density:
\begin{equation}
M_{\rm cloud}=(2\pi)^{3/2} (3/4\pi)^2 (P_{\rm cloud}/G)^{3/2}\rho_{\rm cloud}^{-2}.
\label{mcloud}
\end{equation}
This comes from the equations $P=(\pi/2)G\Sigma^2$, $\Sigma=M/(\pi R^2)$, and
$M=(4\pi/3)\rho R^3$ for average cloud surface density $\Sigma$ and radius $R$.
Setting $M_{\rm cluster}=\epsilon_{\rm M} M_{\rm cloud}$ and $\rho_{\rm
cluster}=\epsilon_{\rm \rho}\rho_{\rm cloud}$ for efficiencies in the gravitating
region equal to $\epsilon_{\rm M}$ and $\epsilon_{\rm \rho}$, and combining these
into a single parameter $\epsilon=\epsilon_{\rm M}\epsilon_{\rm \rho}^2$,
\begin{equation}
M_{\rm cluster}=0.9\epsilon(P_{\rm cloud}/G)^{3/2}\rho_{\rm cluster}^{-2}.
\label{mpr}
\end{equation}
Note that for a given pressure, more massive gravitating regions can form if the
density is lower. Thus OB associations, which have relatively low density, can be
more massive than bound clusters at a given pressure.

Inverting equation (\ref{mpr}),
\begin{equation}
P_{\rm cloud}=G\left(1.1M_{\rm cluster}\rho_{\rm cluster}^2/\epsilon\right)^{2/3}.
\label{pmr}
\end{equation}
For a cluster with $M_{\rm cluster}=10^6\;M_\odot$ inside a half-mass radius of
$\sim3$ pc, the stellar density is $\rho_{\rm
cluster}\sim0.9\times10^4\;M_\odot$ pc$^{-3}$. Setting $\rho_{\rm
cluster}=10^4\;M_\odot$ pc$^{-3}$ and $\epsilon=0.5$ for the core of the cloud,
equation (\ref{pmr}) gives $P_{\rm cloud}\sim8\times10^{10}k_{\rm B}$ for
Boltzman's constant $k_{\rm B}$. For 8 YMCs in \cite{bastian14}, the average
stellar density determined from the ratio of the photometric mass to the volume
at the effective radius is $\sim10^5\;M_\odot$ pc$^{-3}$ and then the average
pressure from equation (\ref{pmr}), considering also the cluster masses, is
$P_{\rm cloud}\sim2\times10^{12}k_{\rm B}$. For comparison, the typical pressure
in the cluster-forming core of a Milky Way cloud, such as the Orion cloud, is
much lower, $5\times10^7$k$_{\rm B}$ from the average of 7 regions observed by
\cite{lada97}. The cluster masses are much lower in Orion too. The average gas
density in these Orion regions is $3\times10^5$ cm$^{-3}$ and the average
one-dimensional velocity dispersion is $\sim0.8$ km s$^{-1}$, which combine to
give the pressure.

Observed cluster densities like that in Orion may not be appropriate for equation
(\ref{pmr}), which assumes a static virialized cloud core.   If gas continuously
streams into a core and forms stars, then the true stellar density can be larger
than $\epsilon_{\rho}$ times the virialized gas density at pressure $P_{\rm
cloud}$. This is the conveyer belt model of \cite{longmore14} and
\cite{walker16}. The effect is to lower the required gas pressure to get a
cluster of a certain stellar density. Effectively, $\epsilon>1$ in this case. For
the following discussion, we assume $P_{\rm cloud}\sim10^{11}k_{\rm B}$ as
representative for $M_{\rm cluster}=10^6\;M_\odot$. Then $\epsilon\sim0.4$ as
above if $\rho_{\rm cluster}\sim10^4\;M_\odot$ pc$^{-3}$; $\epsilon$ increases as
the square of $\rho_{\rm cluster}$ in the conveyer belt model for the same
cluster mass and cloud pressure.

Equation (\ref{pmr}) is based on self-gravitational binding and not feedback. The
role of feedback in limiting the mass of a cluster is not clear.
\cite{ginsburg16} suggest that feedback has virtually no role and gas exhaustion
stops star formation in the cluster, rather than gas expulsion \citep[see
also][]{girichidis12,kruijssen12,matzner17,galvan17,tsang18,silich18,cohen18,ward18}.
Also, there is no jump in the mass function of bound clusters at a mass of around
$10^3\;M_\odot$ where ionization feedback suddenly increases as a result of the
increasingly likely appearance of O-type stars \citep{vacca96}. If gas expulsion
from young stellar feedback was critical in limiting the boundedness of a
cluster, then the probability that a cluster ends up bound should drop suddenly
when feedback from O-type stars spikes upward, causing a comparable drop in the
bound cluster mass function. However this is not observed
\citep[e.g.,][]{bressert10}. Possibly, feedback affects loosely bound clouds
\citep{kruijssen12,matzner15}. Various types of feedback and their relative roles
in cloud core dispersal were discussed in \cite{krumholz18}.

The cluster mass dependence in equation (\ref{pmr}) has not been written fully
yet.  Most likely, cluster radius varies weakly with mass \citep{ryon17} and
cluster density varies more strongly. In Section \ref{convert} we consider that
radius increases approximately as $M^{0.11}$ and $\rho_{\rm cluster}\propto
M^{0.66}$. Then $P_{\rm cloud}\propto M_{\rm cluster}^{1.5}$ from equation
(\ref{pmr}).

\subsection{Conversion from Cloud Core Pressure to Galaxy Gas Surface Density}
\label{conversion}

High pressure in an equilibrium disk requires a high gas surface density, or
actually, a high product of the gas surface density, $\Sigma_{\rm gas}$, and the
total surface density, $\Sigma_{\rm total,GL}$, from the sum of the gas, stars and
dark matter inside the gas layer (subscript 'GL'), which all contribute to the
gravity that weighs down the gas. The resulting pressure is only partly in the
form of turbulent gas motions, which contribute to the core pressure and density
inside a cluster-forming cloud, because other parts of the total pressure are from
magnetic fields and cosmic rays, which resist the weight of the gas too. We let
the turbulent fraction of the pressure be $\zeta\sim0.3$ considering equipartition
with the other types \citep{boulares90}.

For a young galaxy with a high gas fraction, $\Sigma_{\rm gas}\sim0.5\Sigma_{\rm
total,GL}$. For a modern galaxy with a gas disk much thinner than the stellar
disk and the dark matter spheroid, the coefficient is only a little smaller. For
example, in the solar neighborhood, the total stellar midplane density is
$0.043\pm0.04\;M_\odot$ pc$^{-3}$ and the gas density is $1.17$ cm$^{-3}$,
equivalent to $0.041\pm0.004\;M_\odot$ pc$^{-3}$; dark matter adds another
$0.013\pm0.003\;M_\odot$ pc$^{-3}$ \citep{mckee15}. Thus locally, $\Sigma_{\rm
gas}\sim0.4\Sigma_{\rm total,GL}$.  We set $\Sigma_{\rm gas}\sim\xi\Sigma_{\rm
total,GL}$ for the following discussion but because the gas is a little more
concentrated to the midplane than the stars, assume $\xi\sim0.3$ for numerical
evaluations.

The general equation for total pressure in a disk is \citep{elmegreen89},
\begin{equation}
P_{\rm total}=(\pi/2)G\Sigma_{\rm gas}\Sigma_{\rm total,GL}
\label{ptot}
\end{equation}
which comes from $P=\rho\sigma^2(1+\alpha+\beta)$ with gas density
$\rho=\Sigma/(2H)$ and scale height $H=\sigma^2(1+\alpha+\beta)/(\pi G \Sigma_{\rm
total,GL})$ for velocity dispersion $\sigma$, magnetic pressure $\alpha
\rho\sigma^2$, and cosmic ray pressure, $\beta \rho\sigma^2$. Note that $\alpha$
and $\beta$ cancel from equation (\ref{ptot}).

For the cloud pressure, we consider only the turbulent component of the
interstellar pressure, writing $P_{\rm ISM,turb}=\zeta P_{\rm total}$ from above,
and total surface density, $\Sigma_{\rm total,GL}=\Sigma_{\rm gas}/\xi$ to derive
\begin{equation}
P_{\rm ISM,turb}={{\pi\zeta}\over{2\xi}}G\Sigma_{\rm gas}^2.
\label{pturb}
\end{equation}
where $\zeta/\xi\sim1$ (i.e., the additional compression from stars and dark
matter approximately balances the additional expansion from magnetic fields and
cosmic rays).

Equation (\ref{pturb}) is a first step to estimate the minimum gas surface
density in an equilibrium region of a disk that forms a YMC. What we need is the
cloud core pressure, which is used in equation (\ref{pmr}). This core pressure,
$P_{\rm cloud}$, is much larger than the average interstellar turbulent pressure
$P_{\rm ISM,turb}$ because of the additional weight of the cloud envelop pressing
down on the core. We assume these two pressures are proportional to each other
and write $P_{\rm cloud}=CP_{\rm ISM,turb}$ where $C$ is a compaction factor that
could be several orders of magnitude, as estimated below. With this, equations
(\ref{pmr}) and (\ref{pturb}) combine to give
\begin{equation}
\Sigma_{\rm gas}=\left({{2\xi}\over{\pi C\zeta}}\right)^{1/2}
\left({{1.1M_{\rm cluster}\rho_{\rm cluster}^2}\over{\epsilon}}\right)^{1/3}
= 1.0\left({\xi\over{C\zeta}}\right)^{1/2}\epsilon^{-1/3}\Sigma_{\rm cluster}.
\label{sgas}
\end{equation}
Here we have used $M_{\rm cluster}=\Sigma_{\rm cluster}\pi R_{\rm cluster}^2$ and
$\rho_{\rm cluster}=3\Sigma_{\rm cluster}/(4R_{\rm cluster})$ for cluster stellar
surface density $\Sigma_{\rm cluster}$.  With fiducial values of $M_{\rm
cluster}=10^6\;M_\odot$, $\rho_{\rm cluster}=10^4\;M_\odot$ pc$^{-3}$,
$\xi/\zeta=1$, and $\epsilon=0.5$, the average interstellar surface density
should satisfy $\Sigma_{\rm gas}C^{1/2}=4.8\times10^4\;M_\odot$ pc$^{-2}$. This
result is consistent with the peak stellar surface densities in YMCs compiled by
\cite{walker16}, considering the second expression in equation (\ref{sgas}).
\cite{walker16} find that typical YMCs in the Milky Way have $\Sigma_{\rm
cluster}\sim1.3\times10^4\;M_\odot$ pc$^{-2}$.

Now we evaluate $C$. This compaction factor comes from the density stratification
inside molecular clouds. We consider that density varies approximately as a power
law with radius in a spherical cloud, $\rho\propto R^{-\kappa}$, as expected from
self-gravity in either isothermal equilibrium ($\kappa=2$) or collapse
($\kappa=1.5$) conditions \citep{shu77,murray15}. Observations show profiles like
this in individual clouds \citep{mueller02} but they may also be inferred from
the high-density power-law part of the column density probability distribution
functions (N-PDF) in cloud surveys. \cite{corbelli18} determined this
distribution function for molecular clouds in M33 and found in two large regions
an N-PDF slope of around $-2$.  \cite{lin17} observed an N-PDF slope change from
$-4$ to $-2$ in Milky Way dark clouds as the luminosity-to-mass ratio increased.
For active regions of YMC formation, the luminosity-to-mass ratio will be high
and we might expect an N-PDF slope of $\sim-2$. This corresponds to $\kappa\sim2$
using the relation $\kappa=1+2/p$ for N-PDF slope $-p$ \citep{elmegreen18b}. If
cloud internal density varies with radius as $R^{-2}$, then the cloud surface
density varies as $R\rho(R)\propto R^{-1}$, and the cloud pressure, which scales
with the square of the surface density, varies as $(R\rho(R))^2\propto R^{-2}$.
In summary, an N-PDF with a slope of $-p$ on a log-log plot suggests an internal
radial variation of column density inside self-gravitating cloud that has a power
law slope of $-2/p$ and a pressure variation with radius proportional to
$R^{-4/p}$.

This type of dependence makes sense as an approximation if we consider a local
giant molecular cloud like Orion using $p\sim2$.  The average local total ISM
pressure on the scale of the molecular disk thickness, 100 pc \citep{heyer15}, is
$\sim2.8\times10^4k_{\rm B}$ of which about $1/3$ is turbulent \citep{boulares90},
making $P_{\rm ISM,turb}\sim10^4$k$_{\rm B}$. The pressure inside the
cluster-forming core is about a factor of $10^4$ higher, $\sim10^8k_{\rm B}$, as
given above, and the corresponding spatial scale is a factor of $\sim10^2$
smaller, $\sim1$ pc.  Thus average pressure scales about as the inverse square of
size, zooming into the Orion core.

A key assumption here is that molecular clouds like Orion are the densest parts of
a self-gravitating interstellar medium, i.e., with Toomre $Q$ close to 1, and that
the surrounding dark and atomic gas continues to be stratified up to the disk
thickness because this lower-density gas is also part of the total interstellar
gravity \citep{elmegreen87,grabelsky87,elmegreen18}.  This juxtaposition of CO
clouds inside larger HI clouds has been observed in nearby galaxies
\citep{lada88,corbelli18}.

The scaling of pressure with radius inside individual clouds is not easily
evaluated from cloud surveys. Consider the $\sim1100$ CO-emitting clouds in the
\cite{rice16} catalog that have unambiguous distances. The average radius of these
clouds is $34\pm23$ pc, the average log of the surface density (calculated as the
total cloud mass divided by the area) is $1.37\pm0.43$ in units of $M_\odot$
pc$^{2}$ (i.e., the surface density itself is $\sim23\;M_\odot$ pc$^{-2}$), and
the average log of the pressure (calculated as the product of the average density
and the square of the velocity dispersion) is $4.1\pm0.7$ in units of $k_{\rm B}$.
Many of these clouds are not gravitating, however, so their pressures are not
particularly high. For the $\sim400$ clouds that are self-gravitating, with virial
parameter $\alpha<2$, the average log pressure is $4.3\pm0.8$, which is about
twice the ambient turbulent pressure. We should consider also the decrease in
pressure and cloud surface density with galacto-centric radius \citep{heyer15}. If
we just consider the 173 GMCs near the solar position, with galactocentric radii
between 8 and 9 kpc (for the Sun at 8.5 kpc), then the average GMC radius is
$22\pm11$ pc, the average log of the surface density is $1.70\pm0.32$ in $M_\odot$
pc$^{-2}$ ($50\;M_\odot$ pc$^{-2}$), and the average log of the pressure is
$4.7\pm0.5$ in $k_{\rm B}$. These local numbers suggest a scaling of turbulent
cloud pressure with the inverse first power of the size, i.e., the average cloud
is $\sim5$ times smaller than the disk scale height and $\sim5$ times higher in
pressure than average. If we narrowly confine the sample to the solar neighborhood
with low $\alpha$ and a limited mass range, e.g., from $10^4\;M_\odot$ to
$10^5\;M_\odot$, then the pressure scales as approximately the inverse 4th power
of size, following equation (\ref{mcloud}).

A better evaluation of the ratio $C$ of the typical cluster-forming core pressure
to the average interstellar turbulent pressure would seem to come from the
power-law N-PDFs observed by \cite{druard14} and \cite{corbelli18}, which are for
galactic-scale regions in M33, and for Milky Way clouds observed by
\citep{lin17}, which both suggest an average power law slope of around $-2$. This
corresponds to an inverse square radial dependence for density in individual
clouds. Thus it is reasonable to expect that the pressure in a parsec-size cloud
core that forms a YMC is around $C\sim10^4$ times the average ISM turbulent
pressure on the $\sim100$ pc scale of the disk thickness. We assume this value of
$C$ in what follows, even for thicker disks and proportionally bigger cloud cores
at high redshift, because the higher turbulent speeds then \citep{forster09}
should increase the sizes of all self-gravitating regions in proportion to each
other. Now equation (\ref{sgas}) becomes
\begin{equation}
\Sigma_{\rm gas}\sim8.2\times10^{-3}
\left({{M_{\rm cluster}\rho_{\rm cluster}^2}\over{\epsilon}}\right)^{1/3}
= 0.01\epsilon^{-1/3}\Sigma_{\rm cluster}.
\label{sgas2}
\end{equation}
where we have also set $\xi/\zeta=1$.  For the fiducial values used above, i.e.,
$M_{\rm cluster}=10^6\;M_\odot$, $\rho_{\rm cluster}=10^4\;M_\odot$ pc$^{-3}$,
and $\epsilon=0.5$, the minimum interstellar surface density is $\Sigma_{\rm
gas}\sim480\;M_\odot$ pc$^{-2}$. Also for these values, using the second part of
equation (\ref{sgas2}), $\Sigma_{\rm cluster}\sim 3.8\times10^4\;M_\odot$
pc$^{-2}$.  This cluster surface density is smaller by a factor of $\sim3$
compared to the values compiled by \cite{tan14} for local $10^6\;M_\odot$
clusters; a slightly higher cluster density, $4\times10^4\;M_\odot$ pc$^{-3}$
would make them agree better. Then we would derive $\Sigma_{\rm gas}\sim
1300\;M_\odot$ pc$^{-2}$ from the first part of the equation.

\subsection{Conversion from Galaxy Gas Surface Density to Star Formation Rate
Density} \label{convert}

The Kennicutt-Schmidt relation between star formation surface density,
$\Sigma_{\rm SFR}$, and total gas surface density, $\Sigma_{\rm gas}$, may be
written approximately as
\begin{equation} {{\Sigma_{\rm SFR}}\over{M_\odot\;{\rm pc}^{-2}\;{\rm
Myr}^{-1}}}= 0.9\times10^{-4}\left({{\Sigma_{\rm gas}}\over{M_\odot\;{\rm
pc}^{-2}}}\right)^{1.5}. \label{eq:total0}
\end{equation}
for a wide range of $\Sigma_{\rm gas}$ in normal galaxies, i.e., from
$10\;M_\odot$ pc$^{-2}$ to at least $10^4\;M_\odot$ pc$^{-2}$. This expression
was shown by \cite{elmegreen15} to fit the observations in \cite{kennicutt12} for
total galaxy disks and it also fits the set of observations in \cite{krumholz12}
for the same range of $\Sigma_{\rm gas}$ (i.e., \cite{krumholz12} determined a
coefficient of $1.9\times10^{-4}$ and a power of 1.31 including galaxies at
intermediate redshifts). For local galaxies, equation (\ref{eq:total0}) is
consistent with the assumption that the main disks of galaxies evolve toward star
formation on a dynamical time with a constant 1\% efficiency and an approximately
constant gas disk thickness, as observed for molecules in the Milky Way
\citep{heyer15}. The expression may be derived from the dynamical model beginning
with $\Sigma_{\rm SFR}=\epsilon_{\rm ff}\Sigma_{\rm gas}/t_{\rm ff}$, which
converts to
\begin{equation}\Sigma_{\rm SFR}=\epsilon_{\rm ff}(16G/[3\pi H])^{1/2}\Sigma_{\rm
gas}^{3/2}\label{kstotal}
\end{equation}
for $t_{\rm ff}=(32G\rho/[3\pi])^{-1/2}$ and midplane density $\rho=\Sigma/2H$.
The numerical value in equation (\ref{eq:total0}) comes from setting $H=100$ pc
and $\epsilon_{\rm ff}=0.01$ \citep{elmegreen15,elmegreen18b}.

\cite{krumholz12} derived an expression like equation (\ref{eq:total0}) by
assuming the same $\epsilon_{\rm ff}=0.01$ but with $Q\sim1$ for similar-size
galaxies instead of a constant thickness as in equation (\ref{kstotal}). Then the
dynamical time is proportional to the galaxy rotation time. When plotting the
correlation directly in terms of this time, they got
\begin{equation} {{\Sigma_{\rm SFR}}\over{M_\odot\;{\rm pc}^{-2}\;{\rm
Myr}^{-1}}}= 0.22\left({{\Sigma_{\rm gas}}\over{M_\odot\;{\rm
pc}^{-2}}}\right)\left({{T_{\rm orb}}\over{{\rm Myr}}}\right)^{-1},
\label{eq:krum12}
\end{equation}
where $T_{\rm orb}$ is the orbit time at the characteristic radius of the
star-forming gas. Considering that GCs formed at high redshift where galaxy disks
could have been thicker than 100 pc \citep{elmegreen17b}, the empirical relation
in \cite{krumholz12} may be better for GCs than equation (\ref{kstotal}). The
latter may still work if $\epsilon_{\rm ff}$ is larger by a factor of $\sim3$ to
account for the factor of $\sim10$ larger disk thickness, $H$. Both equations
will be used in what follows.

Equation (\ref{kstotal}) allows us to convert the minimum gas surface density for
YMC formation, which comes from the pressure requirement in equation
(\ref{sgas}), to a minimum star formation rate. Substituting $\Sigma_{\rm gas}$
from equation (\ref{sgas}), we obtain
\begin{equation}
\Sigma_{\rm SFR}=\epsilon_{\rm ff}\left({{18GM_{\rm cluster}\rho_{\rm cluster}^2}
\over{3\pi H\epsilon}}\right)^{1/2}
\left({{2\xi}\over{\pi C\zeta}}\right)^{3/4}.
\label{eighteen}
\end{equation}
For $\epsilon_{\rm ff}=0.01$, $H=100$ pc, $\xi/\zeta=1$, $C=10^4$, and
$\epsilon=0.5$ in local galaxies as discussed above,
\begin{equation}
{{\Sigma_{\rm SFR}}\over{M_\odot\;{\rm pc}^{-2}\;{\rm
Myr}^{-1}}}= 0.6\left({{M_{\rm cluster}}\over{10^6\;M_\odot}}\right)^{1/2}
\left({{\rho_{\rm cluster}}\over{10^4\;M_\odot\;{\rm pc}^{-3}}}\right).
\label{point6}
\end{equation}

We can also use the second form of equation (\ref{sgas}) along with equation
(\ref{kstotal}) to obtain
\begin{equation}
\Sigma_{\rm SFR}=\epsilon_{\rm ff}\left({{16G}\over{3\pi H}}\right)^{1/2}
\left({{\xi}\over{C\zeta}}\right)^{3/4}
\epsilon^{-1/2}\Sigma_{\rm cluster}^{3/2}
\label{sixteen}
\end{equation}
and with the same fiducial parameters, this is
\begin{equation}
{{\Sigma_{\rm SFR}}\over{M_\odot\;{\rm pc}^{-2}\;{\rm Myr}^{-1}}}
= 0.083\left({{\Sigma_{\rm cluster}}\over{10^4\;M_\odot\;{\rm pc}^{-2}}}\right)^{3/2}.
\label{point083}
\end{equation}
For typical values appropriate to YMC formation, the minimum $\Sigma_{\rm SFR}$
is $0.1$ to 1 $M_\odot$ pc$^{-2}$ Myr$^{-1}$. This is 35 to 350 times larger than
the value for regions typical of the solar neighborhood, where $\Sigma_{\rm
gas}\sim10\;M_\odot$ and $\Sigma_{\rm SFR}\sim0.003\;M_\odot$ pc$^{-2}$
Myr$^{-1}$.

In a similar way, equations (\ref{eq:krum12}) and (\ref{sgas}) may be combined to
give
\begin{equation}
{{\Sigma_{\rm SFR}}\over{M_\odot\;{\rm pc}^{-2}\;{\rm
Myr}^{-1}}}= 1.1\left({{M_{\rm cluster}}\over{10^6\;M_\odot}}\right)^{1/3}
\left({{\rho_{\rm cluster}}\over{10^4\;M_\odot\;{\rm pc}^{-3}}}\right)^{2/3}
\left({{T_{\rm orb}}\over{100\; {\rm Myr}}}\right)^{-1},
\label{krum12b}
\end{equation}
where again we assume $\xi/\zeta\sim1$, $\epsilon\sim0.5$ and $C=10^4$.

We can write equations (\ref{sgas}) and (\ref{kstotal}) in another way and derive
the maximum likely cluster mass from pressure considerations as a function of the
star formation surface density in main galaxy disks:
\begin{equation}
M_{\rm cluster}={{\epsilon}\over{1.1\rho_{\rm cluster}^2
\epsilon_{\rm ff}^2}} \left({{\pi C \zeta}\over{2\xi}}\right)^{3/2}
\left({{3\pi H}\over{16G}}\right)\Sigma_{\rm SFR}^2.
\label{maindisks}
\end{equation}
For $\epsilon=0.5$, $\epsilon_{\rm ff}=0.01$, $C=10^4$, $\zeta/\xi=1$, and
$H=100$ pc as above,
\begin{equation}
M_{\rm cluster}=1.1\times10^6\;M_\odot\; \left({{\rho_{\rm cluster}}\over{10^4\;M_\odot\;{\rm
pc}^{-3}}}\right)^{-2} \left({{\Sigma_{\rm SFR}}\over{M_\odot\;{\rm pc}^{-2}\;{\rm
Myr}^{-1}}} \right)^2.
\label{mclustersigma}
\end{equation}
Also inverting equation (\ref{krum12b}), we obtain
\begin{equation}
M_{\rm cluster}=0.9\times10^6\;M_\odot\; \left({{\rho_{\rm cluster}}\over{10^4\;M_\odot\;{\rm
pc}^{-3}}}\right)^{-2} \left({{T_{\rm orb}}\over{100\; {\rm Myr}}}\right)^{3}
\left({{\Sigma_{\rm SFR}}\over{M_\odot\;{\rm pc}^{-2}\;{\rm Myr}^{-1}}} \right)^3.
\label{mclustersigmakrum}
\end{equation}

The ratio of the lower limit to the star formation rate from size-of-sample
effects, $\sim1\;M_\odot$ pc$^{-2}$,  discussed in the previous section, and this
lower limit to the star formation rate density from pressure considerations,
$0.1-1\;M_\odot$ pc$^{-2}$ Myr$^{-1}$, gives a characteristic size for
star-forming regions that form massive dense clusters. This size is on the order
of one or a few kiloparsecs for any galaxy large enough to contain at least one of
them. A check on this result is that the duration of star formation should be
several dynamical crossing times on this scale, which, for typical turbulent
speeds of several tens of km s$^{-1}$, is equal to several tens of Myr. This,
combined with the limiting star formation rate, gives a mass of newborn stars
equal to several tens of millions of solar masses, enough to produce a
$\sim10^6\;M_\odot$ cluster and the associated stars. Smaller regions can form GCs
if their star formation surface densities are larger than this limit, as long as
they exceed the total star formation rate.

Equations (\ref{maindisks}) and (\ref{mclustersigma}) have a dependence on the
cluster density as $\rho_{\rm cluster}^{-2}$, which generally depends on cluster
mass.  \cite{ryon17} measured cluster radii for hundreds of clusters in two spiral
galaxies. Combining their results for the galaxies NGC 628 and NGC 1313 and
considering clusters younger than 200 Myr and with masses between
$5\times10^3\;M_\odot$ and $10^5\;M_\odot$, there are 358 clusters that give a
radius-mass relation of
\begin{equation}
\log(R_{\rm cluster})=(0.016 \pm 0.274) + (0.11\pm 0.07)\log(M_{\rm cluster})
\label{ryon1}
\end{equation}
and a density-mass relation of
\begin{equation}
\log(\rho_{\rm cluster})=(-0.670 \pm 0.823) + (0.66\pm 0.20)\log(M_{\rm cluster}).
\label{densitymass}
\end{equation}

\begin{figure*}
\epsscale{1.}
\plotone{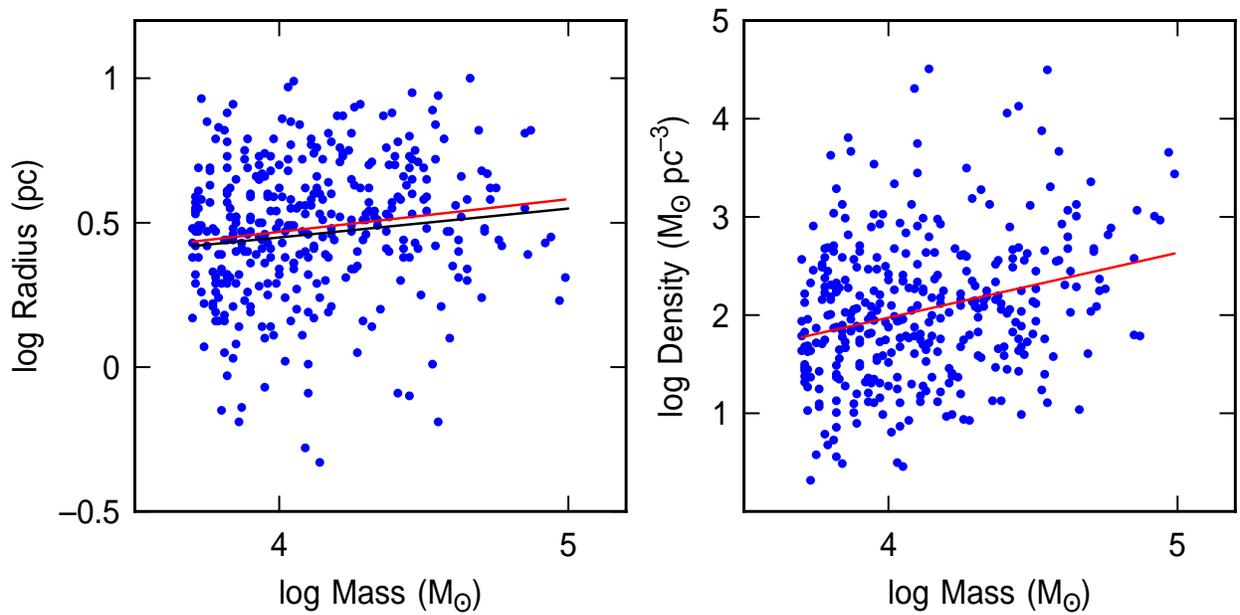}
\caption{(Left) The radii and masses of clusters in two galaxies studied by \cite{ryon17}
and the power-law fit to the small correlation between them. The black line is the
fit from \cite{larsen04}. (Right) The density versus mass of clusters in
Ryon et al., showing the power law fit between them. The fits are given by
equations \ref{ryon1} and \ref{densitymass}.}
\label{ryon_size_mass_new}
\end{figure*}

The values from \cite{ryon17} are plotted in Figure \ref{ryon_size_mass_new} with
densities derived here, and the above fits are shown as red lines. There is a lot
of scatter but a slight trend of radius with mass is evident. \cite{larsen04} also
determined a mass-size relation and got for 18 spiral galaxies $R=AM^B$ with
$A=1.12\pm0.35$ pc and $B=0.10\pm0.03$. That fit is plotted on the left also, as a
black line; it is almost identical to the fit using the data in \cite{ryon17}.

Re-writing equation (\ref{densitymass}),
\begin{equation}
\left({{\rho_{\rm cluster}}\over{10^4\;M_\odot\;{\rm pc}^{-3}}}\right)=
0.20\left({{M_{\rm cluster}}\over{10^6\;M_\odot}}\right)^{0.66},
\label{rhocluster}
\end{equation}
which agrees with typical cluster densities in \cite{port10}. This relation is
assumed to be present at the time of cluster formation, rather than the result of
a long evolution of cluster interactions \citep[e.g.,][]{gieles16}.

Substituting this density dependence into equation (\ref{mclustersigma}) gives
\begin{equation}
\left({{M_{\rm cluster}}\over{10^6\;M_\odot}}\right)^{2.32}=28
\left({{\Sigma_{\rm SFR}}\over{M_\odot\;{\rm pc}^{-2}\;{\rm
Myr}^{-1}}} \right)^2
\label{mclustersigma2}
\end{equation}
or
\begin{equation}
M_{\rm cluster}=4.2\times10^6\;M_\odot \left({{\Sigma_{\rm SFR}}\over{M_\odot\;{\rm pc}^{-2}\;{\rm
Myr}^{-1}}} \right)^{0.86}.
\label{mclustersigma3}
\end{equation}
Similarly, equations (\ref{rhocluster}) and (\ref{mclustersigmakrum}) combine to
give
\begin{equation}
M_{\rm cluster}=3.9\times10^6\;M_\odot \left({{T_{\rm orb}}\over{100\; {\rm Myr}}}\right)^{1.3}
\left({{\Sigma_{\rm SFR}}\over{M_\odot\;{\rm pc}^{-2}\;{\rm Myr}^{-1}}} \right)^{1.3}.
\label{mclustersigma4}
\end{equation}

\cite{johnson17} suggest on the basis of 4 galaxies that the cut-off mass for the
Schechter function is given by $\log(M_{\rm
c})=(6.82\pm0.20)+(1.07\pm0.10)\log(\Sigma_{\rm SFR})$ in the usual units. For
the Antennae galaxy in their data, where $\log(\Sigma_{\rm SFR})=-0.5$, they get
$\log(M_{\rm c})=6.29$ and we get 6.19 from equation (\ref{mclustersigma3}). For
M83 and M51, where $\log(\Sigma_{\rm SFR})\sim-1.5$, they get 5.22 and we get
5.33. For M31, where $\log(\Sigma_{\rm SFR})\sim-2.7$, they get 3.93 and we get
4.30. These values are in reasonable agreement considering the approximations
used for the assumed values of quantities in the pressure and star formation
equations here. A direct comparison is in figure \ref{gcform18_johnson}. The
agreement could be made a little better by fine-tuning some of the assumed
parameters, which are only rough estimates in this paper.  Also shown in Figure
\ref{gcform18_johnson} is equation (\ref{mclustersigma4}) for $T_{\rm orb}=100$
Myr. This equation is not as good a fit, but it does not account for a possible
variation in $T_{\rm orb}$ with $\Sigma_{\rm SFR}$.

\section{The Size-of-Sample Mass compared to the Pressure-Limited Mass}
\label{compare}

For a given size and duration of a star-forming region, sections \ref{minimum} and
\ref{conversion} suggest there is a relationship between the maximum expected mass
of a cluster from the size-of-sample (SOS) effect (eq. \ref{sos}) and the maximum
expected mass from the interstellar pressure (eq. \ref{mclustersigma3}). If the
SOS mass is less than the pressure mass, then the pressure limit is not likely to
be observed because cluster masses will not be sampled far enough into the tail of
the mass function to reach the pressure mass. A similar point was made by
\cite{gieles06b}.  The pressure limit may be related to the cut-off mass, $M_{\rm
c}$, in the Schechter function, or some other function with a cut-off, in which
case the observation of a turn-over in this function depends on the relative
values of these two masses.

In the present section, we identify the cluster mass derived from the cloud-core
pressure and star formation surface density, as given in Section
\ref{sect:mcluster}, with the cut-off mass in the cluster mass distribution
function. It is convenient to write equations (\ref{sos}) and
(\ref{mclustersigma3}) in normalized form:
\begin{equation}
M_{\rm SOS,6}=2 S\Delta t_8
\label{eq:26}
\end{equation}
\begin{equation}
M_{\rm c,6}=4.2\Sigma_{\rm SFR}^{0.86}
\label{eq:27}
\end{equation}
where mass with subscript 6 means in units of $10^6\;M_\odot$, $S$ is the star
formation rate in $M_\odot$ yr$^{-1}$, $\Delta t_8$ is the duration of star
formation in a typical large-scale region in units of $10^8$ yr, and $\Sigma_{\rm
SFR}$ is in units of $M_\odot$ kpc$^{-2}$ yr$^{-1}$ (which is the same as
$M_\odot$ pc$^{-2}$ Myr$^{-1}$).

The cutoff mass is observable in the cluster mass function if
\begin{equation}
{{M_{\rm SOS,6}}\over{M_{\rm c,6}}}>1,
\end{equation}
which is
\begin{equation}
S^{0.14}A^{0.86}\Delta t_8 > 2.1, \;\; {\rm or}\;\; \Sigma_{\rm SFR}^{0.14}A\Delta t_8>2.1
\end{equation}
where we have written $\Sigma_{\rm SFR}=S/A$ for region area $A$ in kpc$^2$. This
result hardly depends on the star formation rates $S$ or $\Sigma_{\rm SFR}$ but
depends mostly on the size and duration of the star formation event, both of which
have to be large, i.e. galaxy-scale, to see the cut-off mass. Even then, the
maximum mass expected stochastically for a region, $M_{\rm SOS}$, is close to the
maximum mass expected from the region pressure, i.e., the cut-off mass, which
implies that the Schechter function form with the cut-off fully sampled should not
be seen clearly in normal galaxies. In fact, it is barely perceptible in the
differential mass functions shown by \cite{adamo17} and \cite{messa17}. The
cut-off will be seen best in large galaxies (high $A$) with long durations of star
formation ($\Delta t$) in each region.

For a given $\Delta t$, equations (\ref{eq:26}) and (\ref{eq:27}) suggest more
simply that the cut-off mass is best seen in galaxies with high star formation
rates and low rate densities, which means large, fairly inactive, galaxies. Large,
low-surface brightness galaxies or large galaxies with fairly low $\Sigma_{\rm
SFR}$ are good examples where the cut-off mass in the Schechter function should be
observable according to these conditions. This is because the pressure is low in
these galaxies, so the maximum cluster mass should be low, but there could be a
lot of clusters in the large disk, sampling far out in the mass function tail.

\section{Comparison to Observations at high redshift}
\label{clump}

Star-forming regions in high-redshift galaxies often exceed the two limits
discussed above, suggesting they commonly make $10^6\;M_\odot$ YMCs. This is
consistent with the discussion in the Introduction which offered many examples
where GC formation was pervasive in the early universe.

\cite{guo18} measured star formation rates in over 3000 clumps in $\sim1000$
galaxies at redshifts from 0.5 to 3. Figure \ref{gcform18_3} shows the
distribution functions of these rates sorted by redshift. The red vertical line in
each panel is the $1\;M_\odot$ yr$^{-1}$ threshold for YMC formation from Section
\ref{minimum}. Background subtraction by the fiducial method was used. The clump
SFRs increase with redshift because of the selection of brighter and more massive
galaxies at greater distances.  The decrease in SFR at low SFR is probably a
selection effect too, since fainter regions are more difficult to see. At the
high-mass end, the distribution functions fall off approximately as power laws.
The three slopes are $-0.85\pm0.06$ for $z=0.5-1$, $-0.85\pm0.06$ for $z=1-2$ and
$-0.68\pm0.10$ for $z=2-3$. These are close to the slope of $\sim-1$ (on a log-log
plot) of the mass functions of clusters and OB associations locally
\citep{port10}, although Figure \ref{gcform18_3} has SFR instead of mass.  Clump
mass functions with this slope were obtained by \cite{dessauges18} using a
different sample of high redshift clumps. For a $-2$ slope, there is an equal
amount of star formation in each equal interval of the log of the SFR. Since the
$1\;M_\odot$ yr$^{-1}$ limit about equally divides the log SFR scale,
approximately half of the star formation in these clumps is in regions that can
produce a $10^6\;M_\odot$ cluster. In fact, the fraction of the observed clump
star formation with a local clump rate greater than $1\;M_\odot$ yr$^{-1}$ is
0.46, 0.61 and 0.72 for the three redshift bins in the figure, respectively.

The surface density of star formation in the \cite{guo18} clumps cannot be
measured because their resolution at $z\sim1$ is only 1.5 kpc diameter.  However,
dividing the star formation rates by the area at 1.5 kpc diameter, which is 1.8
kpc$^2$, gives a rate density that is also comparable to the above limits, namely
$\sim0.5\;M_\odot$ pc$^{-2}$ Myr$^{-1}$.  This is a lower limit to the rate
density because the regions are probably smaller than the resolution.

Star forming clumps at high redshift have also been measured in lensed systems.
\cite{cava18} resolved clumps down to 30 pc and $10^7\;M_\odot$  in the ``snake''.
They found that 55 clumps represent about half of the total star formation rate of
$30\;M_\odot$ yr$^{-1}$, so the average rate among them is $\sim0.24\;M_\odot$
yr$^{-1}$. The most massive of the blue clumps contains $\sim10^{8.5}\;M_\odot$ of
stars in a radius of $\sim300$ pc. If most of these stars formed in $\sim100$ Myr,
then the two rates are $3\;M_\odot$ yr$^{-1}$ and $\sim10\;M_\odot$ pc$^{-2}$
Myr$^{-1}$, both over the threshold for forming a $10^6\;M_\odot$ YMC.

High spatial resolution is also possible in local galaxies that are analogous to
high redshift galaxies in their clump properties. \cite{overzier09} studied 37
clumps in 30 galaxies with $0.1^{\prime\prime}$ resolution at redshifts between
0.1 and 0.3. The average clump radius was 200-400 pc and the average star
formation rate among all the clumps was $0.95\pm0.46\;M_\odot$ yr$^{-1}$. These
quantities give an average rate density of $\sim3\;M_\odot$ pc$^{-2}$ Myr$^{-1}$.
Both values suggest the likely formation of $10^6\;M_\odot$ clusters according to
the present model.

\section{Uncertainties}

This paper considers a cluster mass function that has a characteristic mass below
which the slope is the canonical $-2$ and above which the number of clusters
drops more rapidly. There have been several explanations for this ``cut-off'' or
``maximum'' mass, including fixed fractions of the Jeans mass \citep{kruijssen14}
or the rotationally stabilized mass \citep{escala08} in a galaxy disk. Some
discussion of early models is in \cite{gieles06b}.  Cosmological simulations by
\cite{li17} get a power-law cluster mass function with an upper cut-off too; in
their models, the upper cut-off mass increases with total star formation rate,
but they do not consider the dependence of this cut-off on the star formation
rate density, as in the present paper.

Here we propose that the maximum cluster mass is primarily related to the
pressure in the core of a cluster-forming cloud according to the virial theorem
with a density equal to some fixed factor times the density of the cluster. This
core pressure is then assumed to be $\sim10^4$ times the interstellar pressure --
considering that self-gravitating clouds typically have an internal density
structure that varies approximately with the second power of the inverse of
radius, and because cloud core radii are about 1\% the size of the disk scale
height. The interstellar pressure is then related to the gas surface density by
the usual equilibrium equation, and the gas surface density is related to the
star formation rate surface density by the Kennicutt-Schmidt relation. The result
is a cluster cut-off mass for local galaxies that scales with the 0.86 power of
the star formation rate surface density, and with a normalization that agrees
with observations. Feedback is assumed to play little role in this maximum
cluster mass because the mass is typically so large that the stellar IMF is fully
sampled and the luminosity-to-mass ratio of the stellar mixture has no feature or
sudden increase that might lead to excess disruption of more massive clusters.

Most of the assumptions involved seem to have only a small influence on the
results. A minor role is played by the assumption that the magnetic and cosmic
ray pressures in a galaxy disk nearly balance the stellar and dark matter
contributions to disk self-gravity ($\xi/\zeta\sim1$); this ratio of parameters
is not likely to deviate much from unity, and it agrees with local observations.
Secondly, the ratio of the stellar mass to the virial mass in a cluster-forming
core, or stellar density to virial density, was taken to be represented by the
composite efficiency fraction $\epsilon\sim0.5$. For the so-called conveyer-belt
model of cluster formation (Sect. \ref{sect:minpres}), this dimensionless
parameter could be larger than 1, but it is not likely to be much smaller than
0.5 to get a bound cluster.  Cosmological simulations
\citep[e.g.][]{kimm16,ricotti16,kim18} get high cluster-formation efficiencies
from direct collapse, also supporting a value of $\epsilon$ within a factor of 2
of $\sim0.5$.  This parameter enters $M_{\rm c}$ as $\epsilon^{0.43}$ from
equations (\ref{maindisks}) and (\ref{mclustersigma2}), so the influence on the
result is small.  The efficiency per unit free fall time and disk scale height
for star formation on a galactic scale in the Kennicutt-Schmidt relation were
assumed to be $\epsilon_{\rm ff}\sim0.01$ and $H\sim100$ pc, but these are
important only if the KS law is derived from first principles; if the empirical
law is used, which is reproduced by these values, then this assumption is not
needed. Thus we consider this to be a weak influence on the result also. Another
component is the observed relation between the cluster density and mass, where we
used observations from \cite{ryon17}, reduced to equations (\ref{densitymass})
and (\ref{rhocluster}), and which are also in agreement with a similar
measurement by \cite{larsen04}. Because this is tied to observations, we consider
the uncertainty to be small, but the scatter in the relation is large and the
physical origin of it is unknown.

A more important role is played by the compaction factor $C\sim10^4$, which, as
mentioned above, is the ratio of the pressure in a cluster-forming cloud core to
the ambient pressure in the interstellar medium. This was estimated from
observations in Section \ref{conversion}, so it is not totally unknown, but it
could vary with environment in unknown ways. According to equations
(\ref{maindisks}) and (\ref{mclustersigma2}), it enters into $M_{\rm c}$ as
$C^{0.64}$.  The value of $10^4$ for a cloud density profile that varies as the
inverse square of radius is equivalent to a cluster-forming core with 1\% of the
total cloud mass, giving our pressure-based theory some connection to the
mass-based theories in \cite{kruijssen14} and \cite{escala08}.

Ultimately, the relation between $M_{\rm c}$ and $\Sigma_{\rm SFR}$ will be
measured for many galaxies, and there will also be better measurements of $C$ and
the other parameters suggested here to be important. Then the basic model where
maximum cluster mass depends primarily on interstellar pressure can be checked
more thoroughly.

\section{Conclusions}

The history of star formation in the universe and the build-up of stellar mass
with time both suggest that before a redshift of $\sim2$, which is the time of
peak universal star formation, most regions that formed stars also formed GCs and
their associated unbound stars.  Today, only the most extreme regions of star
formation form GCs.  The transition that must have occurred in cluster-forming
gas is proposed to be the result of a decrease in both the total star formation
rate and the rate per unit area for each independent region, with threshold
values of $\sim1\;M_\odot$ yr$^{-1}$ and $\sim1\;M_\odot$ pc$^{-2}$ Myr$^{-1}$,
respectively, more commonly exceeded in the early universe.

\begin{figure*}
\epsscale{.7}
\plotone{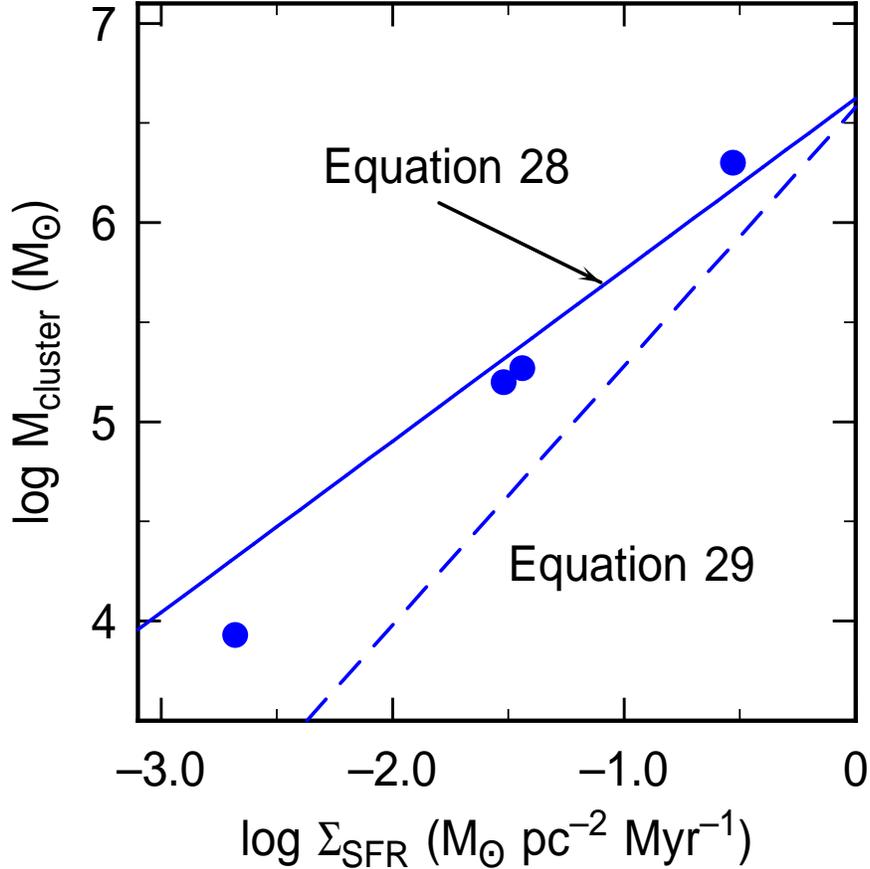}
\caption{The maximum cluster mass as determined from the ISM pressure, cloud
virial equilibrium, and observed cluster density, with a compaction factor of
$C=10^4$ to convert ISM pressure to cloud core pressure, plotted versus
the star formation rate density, which also correlates with the ISM pressure
through the average gas surface density.  The plotted points are from
\cite{johnson17} for the cut-off mass in four galaxies: M31 at low
$\Sigma_{\rm SFR}$, M83 and M51 at intermediate values and the Antennae
at the high value. We identify the maximum cluster mass as determined from
ISM pressure with the cut-off mass in the cluster mass function.  The solid
line uses the Kennicutt-Schmidt relation with a $\Sigma_{\rm gas}^{1.5}$ dependence
tied to observations at low redshift but also approximately applicable at
intermediate redshifts, and the dashed line uses the form of the
Kennicutt-Schmidt relation where the star formation rate density is taken to
scale with the inverse of the orbit time. The latter fit is not as good, but
the orbit time is probably not constant with $\Sigma_{\rm SFR}$ as assumed here;
variation of $T_{\rm orb}$ with the inverse square root of the gas surface density would make
these two lines parallel.}
\label{gcform18_johnson}
\end{figure*}

The maximum mass of a cluster with a characteristic density was linked to the
star formation surface density using the virial equation for cloud core pressure,
a compaction factor that links cloud cores to average interstellar pressures, an
equilibrium equation that associates interstellar pressure with gas surface
density on large-scales, and the Kennicutt-Schmidt relation between gas surface
density and star formation rate surface density. There are many links in this
chain, but each has an observational basis, and the resultant maximum mass agrees
with the observed cluster cut-off mass for a wide range of star formation rate
densities (Fig. \ref{gcform18_johnson}).

\begin{figure*}
\epsscale{1.}
\plotone{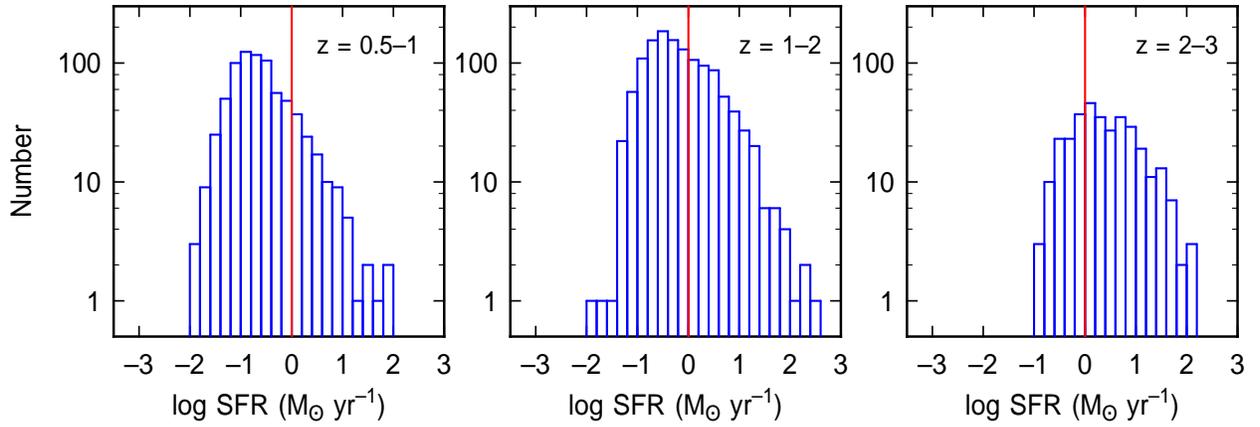}
\caption{Distribution function for star formation rates in the giant clumps of
galaxies studied by \cite{guo18}, divided into three redshift bins. The suggested lower limit for the
formation of a $10^6\;M_\odot$ cluster by the size-of-sample effect
is indicated by the red vertical line. Approximately half of the star formation
in the measured clumps occurs in active-enough clumps to sample a
$10^6\;M_\odot$ cluster. Because these regions are also typically smaller than a
kiloparsec, the surface density of star formation also exceeds the limit for a
$10^6\;M_\odot$ cluster to the right of the red line. }
\label{gcform18_3}
\end{figure*}

The threshold star formation rate and rate density derived here also agree with
observed values in high redshift clumps (Fig. \ref{gcform18_3}), confirming that
these common regions could be the formation sites for most of today's GCs, with
lower-mass galaxies forming the more metal-poor GCs.  The rate density for GC
formation corresponds to a gas surface density of several hundred to a thousand
$M_\odot$ pc$^{-2}$, which is much larger than in today's galaxies but was common
at high redshift.

{\it Acknowledgement:} I am grateful to the referee for useful comments.

\newpage

\end{document}